\documentclass[english]{article}
\usepackage[T1]{fontenc}
\usepackage[latin9]{inputenc}
\usepackage[letterpaper]{geometry}
\usepackage{array}
\usepackage{textcomp}

\makeatletter

\providecommand{\tabularnewline}{\\}

\@ifundefined{definecolor}
 {\usepackage{color}}{}
\makeatother

\makeatother

\usepackage{babel}
\begin{document}

\title{Bidirectional controlled teleportation by using 5-qubit states:\\ A
generalized view}

\maketitle
\begin{center}
Chitra Shukla$^{a}$, Anindita Banerjee$^{b}$ and Anirban Pathak$^{a,c,}$%
\footnote{email: anirban.pathak@gmail.com,

phone: +420 608650694%
}
\par\end{center}

\begin{center}
$^{a}$Department of Physics and Materials Science Engineering,
\par\end{center}

\begin{center}
Jaypee Institute of Information Technology, A-10, Sector-62, Noida,
UP-201307, India.
\par\end{center}

\begin{center}
$^{b}$Department of Physics and Center for Astroparticle Physics
and Space Science,
\par\end{center}

\begin{center}
Bose Institute, Block EN, Sector V, Kolkata 700091, India.
\par\end{center}

\begin{center}
$^{c}$RCPTM, Joint Laboratory of Optics of Palacky University and
Institute of Physics of Academy of Science of the Czech Republic,
Faculty of Science, Palacky University, 17. listopadu 12, 771 46 Olomouc,
Czech Republic.
\par\end{center}

\begin{abstract}
Recently bidirectional controlled perfect teleportation using 5-qubit
states are reported in (Int. J. Theor. Phys. (2013) DOI 10.1007/s10773-013-1484-8
and ibid, (2013) DOI 10.1007/s10773-012-1208-5). In this paper we
have shown that there exists a class of 5-qubit quantum states that
can be used for bidirectional controlled teleportation. Two out of
the three reported cases are the special cases of the proposed class
of 5-qubit quantum states and one of them is not strictly a case of
controlled bidirectional quantum teleportation. Further, we have shown
that one can in principle, construct infinitely many 5-qubit quantum
states for this purpose. We have also shown that the idea can be extended
to bidirectional controlled probabilistic teleportation. Some potential
applications of the proposed scheme and its modified versions are
also discussed in relation with the implementation of quantum remote
control and quantum cryptography.
\end{abstract}
Keywords: bidirectional teleportation, controlled teleportation

\section{Introduction}

Since the introduction of quantum teleportation by Bennett \textit{et
al}. \cite{Bennett} in the year 1993, several modified teleportation
schemes such as quantum information splitting (QIS) or quantum secret
sharing (QSS) \cite{Hillery}, controlled teleportation (CT) \cite{Ct,A.Pathak},
hierarchical quantum information splitting (HQIS) \cite{hierarchical,Shukla},
remote state preparation \cite{Pati} etc. are prescribed. The teleportation
schemes and its modifications draw considerable attention of the quantum
communication community. This is  because they have no classical analogue
and  they are useful for secure quantum communication
and remote quantum operations \cite{J. A. Vaccaro}. The original scheme
of teleportation was a one-way scheme in which Alice sends an unknown
single qubit quantum state to Bob by using two bits of classical communication
and an entangled state already shared by Alice and Bob. Subsequently,
Huelga \textit{et al}. \cite{J. A. Vaccaro,J. A. Vaccaro-1} and others
discussed the possibility of using bidirectional state teleportation
(BST) for the implementation of nonlocal quantum gates. In the schemes
proposed by Huelga \textit{et al}. sharing of entanglement and classical
transmission are the required resources. These resources were quantified
by them and a lower bound on the resources required
to perform quantum remote control (i.e. teleportation of an arbitrary
unitary operation $U$ ) was established.

Recently Zha \textit{et al}. \cite{Zha,Zha II} and Li \textit{et
al}. \cite{Li} have reported tripartite schemes for bidirectional
controlled teleportation (BCST). Although Zha \textit{et al}. and
Li \textit{et al}. have not illustrated their schemes as a generalization
of BST, it is easy to recognize a BCST scheme as
a generalization of BST scheme. To be precise, a BCST scheme is a
three party scheme where BST is possible provided the supervisor/controller
(Charlie) discloses his information. It is important to note that
the control of supervisor Charlie should be in both the direction
of communication.

In a BST scheme Alice and Bob can simultaneously send an unknown quantum
states to each other. The usefulness of BST can be understood clearly
if we consider a simple scenario in which Bob teleports a single qubit
state $|\psi\rangle$ to Alice who applies an unitary operator $U$
on $|\psi\rangle$ and teleports back the state $|\psi^{\prime}\rangle=U|\psi\rangle$
to Bob. The above described scenario is nothing but BST but we can
quickly recognize that it can be used to implement a nonlocal quantum
gate or a quantum remote control. We may now think of a situation
where Charlie is boss and Alice and Bob are his subordinates who are
semihonest. For a specific task Alice and Bob may require to implement
a quantum remote control. However, Alice and Bob are allowed to implement
the quantum remote control only when Charlie permits them to do so.
In such a scenario we need schemes for BCST. This is clearly a special
case of a quantum cryptographic switch recently introduced by one
of the present authors \cite{switch}. These observations have motivated
us to investigate the intrinsic symmetry of the 5-qubit quantum states
that are useful for the implementation of BCST.

The remaining part of the paper is organized as follows. In Section
\ref{sec:General-structure-of} we have described a set of quantum states
that may be used for BCST and have shown that the recently reported
works \cite{Zha,Zha II} are special cases of a more general scheme.
It is also shown that the control of Charlie on the BCST scheme reported
by Li \textit{et al}. is limited to one direction only. A notion of
probabilistic BCST is introduced in Section \ref{sec:Probabilistic-bct}
and finally the work is concluded in the Section \ref{sec:Conclusion}.

\section{General structure of the quantum states to be used for Bidirectional
Controlled Teleportation\label{sec:General-structure-of}}

The 5-qubit quantum states that are useful for BCST may be described
as
\begin{equation}
|\psi\rangle_{12345}=\frac{1}{\sqrt{2}}\left(|\psi_{1}\rangle_{A_{1}B_{1}}|\psi_{2}\rangle_{A_{2}B_{2}}|a\rangle_{C_{1}}\pm|\psi_{3}\rangle_{A_{1}B_{1}}|\psi_{4}\rangle_{A_{2}B_{2}}|b\rangle_{C_{1}}\right),\label{eq:the 5-qubit state}
\end{equation}
where single qubit states $|a\rangle$ and $|b\rangle$ satisfy $\langle a|b\rangle=\delta_{a,b}$,
$|\psi_{i}\rangle\in\left\{ |\psi^{+}\rangle,|\psi^{-}\rangle,|\phi^{+}\rangle,|\phi^{-}\rangle:|\psi_{1}\rangle\neq|\psi_{3}\rangle,|\psi_{2}\rangle\neq|\psi_{4}\rangle\right\} $,
$|\psi^{\pm}\rangle=\frac{|00\rangle\pm|11\rangle}{\sqrt{2}},$ $|\phi^{\pm}\rangle=\frac{|01\rangle\pm|10\rangle}{\sqrt{2}}$
and the subscripts $A$, $B$ and $C$ indicate the qubits of Alice,
Bob and Charlie respectively. Thus $|\psi_{i}\rangle$ is a Bell state. The condition
\begin{equation}
|\psi_{1}\rangle\neq|\psi_{3}\rangle,|\psi_{2}\rangle\neq|\psi_{4}\rangle\label{eq:condition}
\end{equation}
ensures that Charlie's qubit is appropriately entangled with remaining
4 qubits. By appropriately entangled we mean that unless Charlie measures
his qubit in $\{|a\rangle,|b\rangle\}$ basis and discloses the outcome.
Alice and Bob are unaware of the entangled (Bell) states they share
and consequently the receiver does not know upon the receipt of the
measurement outcome of the sender which unitary operation is to be
applied. In case $|\psi_{1}\rangle=|\psi_{3}\rangle$ ($|\psi_{2}\rangle=|\psi_{4}\rangle)$
is allowed then the qubits 1 and 2 (3 and 4) are separable from the
remaining qubits and consequently Charlie has no control over the
teleportation done using those two qubits. Now when the state (\ref{eq:the 5-qubit state})
satisfies the condition (\ref{eq:condition}) then on the disclosure
of the outcome of Charlie's measurement on $\{|a\rangle,|b\rangle\}$
basis, Alice and Bob knows with certainty which two Bell states they
share and consequently they can use the conventional teleportation
scheme to teleport unknown quantum states. The notion would be more clear
from the Table \ref{tab:table1}, which clearly shows that without
the knowledge of the initial Bell states shared by Alice and Bob,
the receiver cannot decide the operation to be implemented by him/her.
As the condition (\ref{eq:condition}) ensures that without the disclosure
of Charlie the receiver and the sender do not know the entangled state
shared by them so Charlie has a control over the bidirectional teleportation
scheme.

The quantum state used for BCST by Zha \textit{et al}. \cite{Zha}
is
\begin{equation}
|\psi_{{\rm Zha}}\rangle_{12345}=\frac{1}{2}\left(|00000\rangle+|00111\rangle+|11010\rangle+|11101\rangle\right)_{12345},\label{eq:zha-initial}
\end{equation}
 where the qubits 1 and 3 belong to Alice, qubits 2 and 5 belong to
Bob and the qubit 4 is with Charlie. Now we can rearrange the state
(\ref{eq:zha-initial}) as
\begin{equation}
|\psi_{{\rm Zha}}\rangle_{12354}=\frac{1}{\sqrt{2}}\left(|\psi^{+}\rangle_{12}|\psi^{+}\rangle_{35}|+\rangle_{4}+|\psi^{-}\rangle_{12}|\psi^{-}\rangle_{35}|-\rangle_{4}\right).\label{eq:zha in our form}
\end{equation}
Clearly (\ref{eq:zha in our form}) is in the form (\ref{eq:the 5-qubit state})
and it satisfies the condition (\ref{eq:condition}). Consequently,
$|\psi_{{\rm Zha}}\rangle$ is a special case of the class of state
described by (\ref{eq:the 5-qubit state}), which are helpful for
bidirectional quantum teleportation. To be precise, Alice and Bob
do not know the Bell states they share unless Charlie (supervisor)
discloses the outcome of the measurement performed by him using $\{|+\rangle,|-\rangle\}$
basis. On the other hand, on disclosure of Charlie's outcome Alice
and Bob obtain complete knowledge of Bell state they share and subsequently
they may use Table \ref{tab:table1} for successful teleportation.

Similarly, a BCST scheme was proposed by  Zha \textit{et al}. \cite{Zha II} using modified Brown state which  is

\begin{equation}
|\psi_{{\rm Zha}}^{\prime}\rangle_{12345}=\frac{1}{2\sqrt{2}}\left(-|11101\rangle+|11110\rangle+|00000\rangle-|00011\rangle+|01001\rangle+|01010\rangle+|10100\rangle+|10111\rangle\right)_{12345},\label{eq:initial ZHA II}
\end{equation}
where the qubits 1 and 2 belong to Alice, qubits 3 and 4 belong to
Bob and qubit 5 is with Charlie. Charlie (supervisor) measures in
$\{|0\rangle,|1\rangle\}$ basis. Now we can rearrange the state (\ref{eq:initial ZHA II})
as

\begin{equation}
|\psi_{{\rm Zha}}^{\prime}\rangle_{12354}=\frac{1}{\sqrt{2}}\left(|\psi^{+}\rangle_{13}|\psi^{+}\rangle_{24}|0\rangle_{5}-|\psi^{-}\rangle_{13}|\phi^{-}\rangle_{24}|1\rangle_{5}\right).\label{eq:ZHA II in our form}
\end{equation}
Clearly, (\ref{eq:ZHA II in our form}) is in the form (\ref{eq:the 5-qubit state})
and it satisfies the condition (\ref{eq:condition}). So, $|\psi_{Zha}^{\prime}\rangle$
is also helpful for bidirectional quantum teleportation (Table \ref{tab:table1}
can be used for successful teleportation) and is a special case of
the class of state described by (\ref{eq:the 5-qubit state}).

\begin{table}
\begin{centering}
\begin{tabular}{|c|c|c|c|c|}
\hline
 & \multicolumn{4}{c|}{Initial state shared by Alice and Bob}\tabularnewline
\hline
 & $|\psi^{+}\rangle$ & $|\psi^{-}\rangle$ & $|\phi^{+}\rangle$ & $|\phi^{-}\rangle$\tabularnewline
\hline
SMO & Receiver& Receiver & Receiver & Receiver\tabularnewline
\hline
00 & $I$ & $Z$ & $X$ & $iY$\tabularnewline
\hline
01 & $X$ & $X$ & $I$ & $Z$\tabularnewline
\hline
10 & $Z$ & $I$ & $iY$ & $X$\tabularnewline
\hline
11 & $iY$ & $iY$ & $Z$ & $I$\tabularnewline
\hline
\end{tabular}
\par\end{centering}

\caption{\label{tab:table1} Perfect Teleportation. Here SMO stands for sender's measurement outcome.}
\end{table}

Interestingly, the 5-qubit quantum state used by Li \textit{et al}.
\cite{Li} does not satisfy the condition (\ref{eq:condition}). To
be precise, to implement bidirectional quantum teleportation Li \textit{et
al}. \cite{Li} have used 5-qubit quantum state
\begin{equation}
|\psi_{Li}\rangle_{12345}=\frac{1}{\sqrt{2}}\left(|000\rangle+|111\rangle\right)_{123}\otimes\frac{1}{\sqrt{2}}\left(|00\rangle+|11\rangle\right)_{45}\label{eq:li}
\end{equation}
where qubits 3 and 5 are with Alice, qubits 1 and 4 are with Bob and
the qubit 2 belongs to Charlie. As Charlie keeps only the second qubit
with him, so Alice and Bob have access to rest of the qubits. Clearly
Alice and Bob can use $|\psi^{+}\rangle_{54}=\frac{1}{\sqrt{2}}\left(|00\rangle+|11\rangle\right)_{54}$
to teleport an unknown state without the control of Charlie. Essentially
Charlie has control over the communication in one direction only and
the scheme described by Li \textit{et al}. fails to control the bidirectional
aspect of teleportation. It's not surprising as
\begin{equation}
|\psi_{Li}\rangle=\frac{1}{\sqrt{2}}\left(|\psi^{+}\rangle_{31}|+\rangle_{2}+|\psi^{-}\rangle_{31}|-\rangle_{2}\right)|\psi^{+}\rangle_{54},\label{eq:li-1}
\end{equation}
which does not satisfy the condition (\ref{eq:condition}). Moreover,
it may be noted that the paper of Li \textit{et al}. \cite{Li} unfortunately
contains a few mistakes. For example, in the Eqn. (6) of \cite{Li}
there is a typo of negative sign in the second term $(a_{0}|0\rangle-a_{1}|1\rangle)_{3},$
which should be $(a_{0}|0\rangle+a_{1}|1\rangle)_{3}$. Table 1 of
their paper is correct and prepared according to the corrected Eqn. 
(6) but it is noticeable that when Charlie measures in $\{|+\rangle,|-\rangle\}$
basis whatever be the outcome of his measurement Alice and Bob need
to apply the same unitary operation, for example whether he got $|+\rangle_{2}$
or $|-\rangle_{2}$, in both the situation Alice and Bob need to apply
the same unitary operation $I_{3}\otimes I_{4}$ (see the first two
rows of Table 1 \cite{Li}). Consequently, Charlie does not have the
required control over the BCST scheme. To be precise his control is
limited to one direction only. So the scheme proposed by Li \textit{et
al}. is not that of BCST and as a natural consequence of this observation
we find that the 5-qubit state used by them is not a member of our
set of quantum states described by (\ref{eq:the 5-qubit state}) and
(\ref{eq:condition}).

If we do not restrict us by the condition (\ref{eq:condition}), then
for each choice of basis set $\{|a\rangle,|b\rangle\}$ for the measurement
of Charlie, there exists 256 possible quantum states of the form (\ref{eq:the 5-qubit state})
without considering the relative phase ($\pm$ sign in the middle).
Out of which there are 64 cases where $|\psi_{1}\rangle=|\psi_{3}\rangle$
($|\psi_{2}\rangle=|\psi_{4}\rangle$). Similarly there are 64 cases
where $|\psi_{2}\rangle=|\psi_{4}\rangle$. However, there exist 16
cases where $|\psi_{1}\rangle=|\psi_{3}\rangle$ and $|\psi_{2}\rangle=|\psi_{4}\rangle.$
Thus total number of ways in which we can obtain a 5-qubit state that
can be used for bidirectional teleportation is $256-2\times64+16=144.$
As an example, in Table (\ref{tab:Table 2}) we have shown a subset of possible
choices of $\{|\psi_{i}\rangle\}$ that satisfies condition (\ref{eq:condition}).
Now since $\{|a\rangle,|b\rangle\}$ can be chosen in infinitely many
ways, in principle we can perform bidirectional controlled teleportation
in infinitely many ways by using quantum states of the form (\ref{eq:the 5-qubit state}).
It is obvious and it does not make any sense to further investigate
a particular state using the approach adopted in \cite{Zha} or in
\cite{Zha II}.
\begin{table}
\centering{}%
\begin{tabular}{|c|c|c|c|}
\hline
$|\psi_{1}\rangle$ & $|\psi_{2}\rangle$ & $|\psi_{3}\rangle$ & $|\psi_{4}\rangle$\tabularnewline
\hline
$|\psi^{+}\rangle$ & $|\psi^{+}\rangle$ & $|\phi^{+}\rangle$ & $|\phi^{+}\rangle$\tabularnewline
\hline
$|\psi^{+}\rangle$ & $|\psi^{+}\rangle$ & $|\phi^{+}\rangle$ & $|\phi^{-}\rangle$\tabularnewline
\hline
$|\psi^{+}\rangle$ & $|\psi^{+}\rangle$ & $|\phi^{-}\rangle$ & $|\phi^{-}\rangle$\tabularnewline
\hline
$|\psi^{+}\rangle$ & $|\psi^{+}\rangle$ & $|\phi^{-}\rangle$ & $|\phi^{+}\rangle$\tabularnewline
\hline
$|\psi^{+}\rangle$ & $|\psi^{+}\rangle$ & $|\psi^{-}\rangle$ & $|\psi^{-}\rangle$\tabularnewline
\hline
$|\psi^{+}\rangle$ & $|\psi^{+}\rangle$ & $|\psi^{-}\rangle$ & $|\phi^{-}\rangle$\tabularnewline
\hline
$|\psi^{+}\rangle$ & $|\psi^{+}\rangle$ & $|\phi^{-}\rangle$ & $|\psi^{-}\rangle$\tabularnewline
\hline
$|\psi^{+}\rangle$ & $|\psi^{+}\rangle$ & $|\psi^{-}\rangle$ & $|\phi^{+}\rangle$\tabularnewline
\hline
$|\psi^{+}\rangle$ & $|\psi^{+}\rangle$ & $|\phi^{+}\rangle$ & $|\psi^{-}\rangle$\tabularnewline
\hline
\end{tabular}\caption{\label{tab:Table 2} A subset of possible choices of $\{|\psi_{i}\rangle\}$
that satisfies condition (\ref{eq:condition}) and may be used to
form quantum states of the form (\ref{eq:the 5-qubit state}) which
will be useful for BCST. }
\end{table}

\section{Probabilistic bidirectional controlled teleportation\label{sec:Probabilistic-bct}}

If we wish to extend the idea for probabilistic teleportation $|\psi_{i}\rangle\in\left\{ |\psi^{\prime+}\rangle,|\psi^{\prime-}\rangle,|\phi^{\prime+}\rangle,|\phi^{\prime-}\rangle:|\psi_{1}\rangle\neq|\psi_{3}\rangle,|\psi_{2}\rangle\neq|\psi_{4}\rangle\right\} $,
$|\psi^{\prime\pm}\rangle=a_{1}|00\rangle\pm b_{1}|11\rangle,$ $|\phi^{\prime\pm}\rangle=a_{2}|01\rangle\pm b_{2}|10\rangle$
where $|a_{i}|^{2}+|b_{i}|^{2}=1$ and $|a_{i}|\neq\frac{1}{\sqrt{2}}$.
Now we may follow the usual scheme of teleportation with only difference
that the receiver cannot construct a single qubit unitary operation
to map $\frac{\alpha a_{i}|0\rangle\pm\beta b_{i}|1\rangle}{\sqrt{|\alpha a_{i}|^{2}+|\beta b_{i}|^{2}}}$
to the unknown quantum state $\alpha|0\rangle+\beta|1\rangle$ without
the knowledge of $\alpha$ and $\beta$. For this reason, the receiver
is required to change his/her strategy. He/she has to prepare an ancilla
qubit in $|0\rangle_{Auxi}$ and applies $U$ or $U_{1}$ unitary
operations on the combined system $\frac{\alpha a_{i}|0\rangle\pm\beta b_{i}|1\rangle}{\sqrt{|\alpha a_{i}|^{2}+|\beta b_{i}|^{2}}}|0\rangle_{Auxi}$
(i.e. of his/her existing qubit and ancilla) depending on the initial
state; where

$U=\left(\begin{array}{lclc}
\frac{b}{a} & \sqrt{1-\frac{b^{2}}{a^{2}}} & 0 & 0\\
0 & 0 & 0 & -1\\
0 & 0 & 1 & 0\\
\sqrt{1-\frac{b^{2}}{a^{2}}} & -\frac{b}{a} & 0 & 0
\end{array}\right)$ and $U_{1}=U(X\otimes I)=\left(\begin{array}{lclc}
0 & 0 & \frac{b}{a} & \sqrt{1-\frac{b^{2}}{a^{2}}}\\
0 & -1 & 0 & 0\\
1 & 0 & 0 & 0\\
0 & 0 & \sqrt{1-\frac{b^{2}}{a^{2}}} & -\frac{b}{a}
\end{array}\right)$. Subsequently the receiver can measure his/her qubit (ancilla) in
the computational basis. If the receiver's measurement outcome yields
$|0\rangle$ then he/she obtains unknown state with unit fidelity
but if his/her measurement outcome on ancilla yields $|1\rangle$
then the teleportation fails and will not workout. Now supervisor
discloses his/her outcome of measurement then sender and receiver
would be able to get the complete knowledge of Bell state they share
and subsequently they may use Table \ref{tab:For-Probabilistic-Teleportation} to construct the unknown state teleported by the sender.

\begin{center}
\begin{table}
\begin{centering}
\begin{tabular}{|c|c|c|c|c|}
\hline
 & \multicolumn{4}{c|}{Initial state shared by Alice and Bob}\tabularnewline
\hline
 & $|\psi^{\prime+}\rangle$ & $|\psi^{\prime-}\rangle$ & $|\phi^{\prime+}\rangle$ & $|\phi^{\prime-}\rangle$\tabularnewline
\hline
 & \multicolumn{2}{c|}{When $U$ operation is applied} & \multicolumn{2}{c|}{When $U_{1}$ operation is applied}\tabularnewline
\hline
SMO & Receiver's operation & Receiver's operation & Receiver's operation & Receiver's operation\tabularnewline
\hline
00 & $I$ & $Z$ & $I$ & $Z$\tabularnewline
\hline
01 & $X$ & $iY$ & $X$ & $iY$\tabularnewline
\hline
10 & $Z$ & $I$ & $Z$ & $I$\tabularnewline
\hline
11 & $iY$ & $X$ & $iY$ & $X$\tabularnewline
\hline
\end{tabular}
\par\end{centering}

\caption{\label{tab:For-Probabilistic-Teleportation} Probabilistic Teleportation}
\end{table}

\par\end{center}

\begin{table}
\begin{centering}
\begin{tabular}{|>{\centering}p{1.5in}|c|}
\hline
Initial product state of Alice and Bob (after the measurement of Charlie) & Rearranged state of Alice and Bob\tabularnewline
\hline
$|\psi^{+}\psi^{+}\rangle_{1234}$ & $\frac{1}{2}\left\{ |\psi^{+}\psi^{+}\rangle+|\phi^{+}\phi^{+}\rangle+|\phi^{-}\phi^{-}\rangle+|\psi^{-}\psi^{-}\rangle\right\} _{1324}$\tabularnewline
\hline
$|\psi^{-}\psi^{-}\rangle_{1234}$ & $\frac{1}{2}\left\{ |\psi^{+}\psi^{+}\rangle-|\phi^{+}\phi^{+}\rangle-|\phi^{-}\phi^{-}\rangle+|\psi^{-}\psi^{-}\rangle\right\} _{1324}$\tabularnewline
\hline
$|\psi^{+}\psi^{-}\rangle_{1234}$ & $\frac{1}{2}\left\{ |\psi^{+}\psi^{-}\rangle-|\phi^{+}\phi^{-}\rangle-|\phi^{-}\phi^{+}\rangle+|\psi^{-}\psi^{+}\rangle\right\} _{1324}$\tabularnewline
\hline
$|\psi^{-}\psi^{+}\rangle_{1234}$ & $\frac{1}{2}\left\{ |\psi^{+}\psi^{-}\rangle+|\phi^{+}\phi^{-}\rangle+|\phi^{-}\phi^{+}\rangle+|\psi^{-}\psi^{+}\rangle\right\} _{1324}$\tabularnewline
\hline
$|\psi^{+}\phi^{+}\rangle_{1234}$ & $\frac{1}{2}\left\{ |\psi^{+}\phi^{+}\rangle+|\psi^{-}\phi^{-}\rangle+|\phi^{+}\psi^{+}\rangle+|\phi^{-}\psi^{-}\rangle\right\} _{1324}$\tabularnewline
\hline
$|\psi^{-}\phi^{-}\rangle_{1234}$ & $\frac{1}{2}\left\{ |\psi^{+}\phi^{+}\rangle+|\psi^{-}\phi^{-}\rangle-|\phi^{+}\psi^{+}\rangle-|\phi^{-}\psi^{-}\rangle\right\} _{1324}$\tabularnewline
\hline
$|\phi^{+}\psi^{+}\rangle_{1234}$ & $\frac{1}{2}\left\{ |\psi^{+}\phi^{+}\rangle+|\phi^{+}\psi^{+}\rangle-|\phi^{-}\psi^{-}\rangle-|\psi^{-}\phi^{-}\rangle\right\} _{1324}$\tabularnewline
\hline
$|\phi^{-}\psi^{-}\rangle_{1234}$ & $\frac{1}{2}\left\{ |\psi^{+}\phi^{+}\rangle-|\phi^{+}\psi^{+}\rangle+|\phi^{-}\psi^{-}\rangle-|\psi^{-}\phi^{-}\rangle\right\} _{1324}$\tabularnewline
\hline
$|\psi^{+}\phi^{-}\rangle_{1234}$ & $\frac{1}{2}\left\{ |\psi^{+}\phi^{+}\rangle+|\psi^{-}\phi^{-}\rangle-|\phi^{+}\psi^{-}\rangle-|\phi^{-}\psi^{+}\rangle\right\} _{1324}$\tabularnewline
\hline
$|\psi^{-}\phi^{+}\rangle_{1234}$ & $\frac{1}{2}\left\{ |\psi^{+}\phi^{+}\rangle+|\psi^{-}\phi^{-}\rangle+|\phi^{+}\psi^{-}\rangle+|\phi^{-}\psi^{+}\rangle\right\} _{1324}$\tabularnewline
\hline
$|\phi^{+}\psi^{-}\rangle_{1234}$ & $\frac{1}{2}\left\{ |\psi^{-}\phi^{+}\rangle-|\psi^{+}\phi^{-}\rangle+|\phi^{+}\psi^{-}\rangle-|\phi^{-}\psi^{+}\rangle\right\} _{1324}$\tabularnewline
\hline
$|\phi^{-}\psi^{+}\rangle_{1234}$ & $\frac{1}{2}\left\{ |\psi^{-}\phi^{+}\rangle-|\psi^{+}\phi^{-}\rangle-|\phi^{+}\psi^{-}\rangle+|\phi^{-}\psi^{+}\rangle\right\} _{1324}$\tabularnewline
\hline
$|\phi^{+}\phi^{+}\rangle_{1234}$ & $\frac{1}{2}\left\{ |\psi^{+}\psi^{+}\rangle+|\phi^{+}\phi^{+}\rangle-|\phi^{-}\phi^{-}\rangle-|\psi^{-}\psi^{-}\rangle\right\} _{1324}$\tabularnewline
\hline
$|\phi^{-}\phi^{-}\rangle_{1234}$ & $\frac{1}{2}\left\{ |\psi^{+}\psi^{+}\rangle-|\phi^{+}\phi^{+}\rangle+|\phi^{-}\phi^{-}\rangle-|\psi^{-}\psi^{-}\rangle\right\} _{1324}$\tabularnewline
\hline
$|\phi^{+}\phi^{-}\rangle_{1234}$ & $\frac{1}{2}\left\{ |\phi^{+}\phi^{-}\rangle-|\phi^{-}\phi^{+}\rangle-|\psi^{+}\psi^{-}\rangle+|\psi^{-}\psi^{+}\rangle\right\} _{1324}$\tabularnewline
\hline
$|\phi^{-}\phi^{+}\rangle_{1234}$ & $\frac{1}{2}\left\{ |\phi^{+}\phi^{+}\rangle-|\phi^{-}\phi^{-}\rangle-|\psi^{-}\psi^{-}\rangle+|\psi^{+}\psi^{+}\rangle\right\} _{1324}$\tabularnewline
\hline
\end{tabular}
\par\end{centering}

\caption{\label{tab:ent-swapping}In the left column we show the product state
of Alice and Bob, where qubits 1 and 3 are with Alice and qubits 2 and
4 are with Bob. In the right column the same product state is rearranged.
Now if Alice measures particle 1, 3 in Bell basis then particles 2,
4 will collapse to a Bell state which is uniquely connected to the
outcome of Alice. From the outcome of his own Bell measurement Bob
can conclude the outcome of Alice provided he knows the initial state
he shares with Alice. To know that he needs to know the outcome of
Charlie. }
\end{table}

\section{Conclusions\label{sec:Conclusion}}

A set of schemes of BCST have recently been published using different
5-qubit quantum states like 5-qubit Cluster state \cite{Zha}, 5-qubit
Brown state \cite{Zha II} and 5-qubit composite GHZ-Bell state \cite{Li}
etc. However, the link of BCST with the quantum remote control and
the practical applicability of the schemes were not discussed. In
the present paper we have already described some of the important
applications of the BCST schemes. We may now further note that it's
easy to turn a BST scheme into a LM-05 \cite{LM05} type protocol
of quantum secure direct communication (QSDC) which can be reduced
to a protocol of quantum key distribution (QKD). In a BST type version
of LM-05 protocol the quantum states will be teleported from Alice
to Bob and vice versa. Consequently information encoded quantum states
will not be available in the channel. This would help us to circumvent
different types of eavesdropping attacks. A similar scheme of deterministic
secure quantum communication (DSQC) without the actual transmission
of the key is recently discussed by Zhang \textit{et al}. \cite{ent-swapping}.
They have used entanglement swapping to communicate a secure message.
For entanglement swapping we need a product state which is a product
of two Bell state. Now after the measurement of Charlie the state
of Alice and Bob in the present scheme of BCST is just a product state
of two Bell states. Now Alice performs a Bell measurement on the 2 particles
available with her and notes $00,\,01,\,10$ and $11$ as key if she
obtains $|\psi^{+}\rangle,|\psi^{-}\rangle,|\phi^{+}\rangle,$ and
$|\phi^{-}\rangle$ respectively. Alice does not require to announce
her outcome. Subsequently Bob performs a Bell measurement on his qubits.
His outcome is uniquely related to the outcome of Alice as shown in
the Table \ref{tab:ent-swapping}. However to infer the outcome of
Alice from his own outcome Bob would require to know the outcome of
Charlie. Thus if we consider Alice and Bob as semihonest they will
be able to generate a quantum key using the 5-qubit quantum state
(\ref{eq:the 5-qubit state}) only when the supervisor Charlie allows
them to do so. The significance of BCST discussed here was not discussed
in the earlier works \cite{Zha,Zha II,Li}. However, identification
of its practical applicability makes it a more relevant and motivating
problem to explore. Further we have established that there exists
a set of quantum states which can be used for BCST and the states
used by Zha \textit{et al}. in \cite{Zha} and \cite{Zha II} are
of the elements of that set. The identification of a large set of
quantum states that are useful for BCST has increased the possibility
of experimental realization of BCST. Keeping this in mind we end this
short paper with the expectation that several new application of BCST
will be found in near future and experimental realization of BCST
and its applications will also be possible in near future.

Acknowledgment: AP thanks Department of Science and Technology (DST),
India for support provided through the DST project No. SR/S2/LOP-0012/2010
and he also acknowledges the supports received from the projects CZ.1.05/2.1.00/03.0058
and CZ.1.07/2.3.00/20.0017 of the Ministry of Education, Youth and
Sports of the Czech Republic.


\end{document}